\documentclass[12pt, preprint]{aastex}        % standard for submission

\slugcomment{The Astrophysical Journal Letters, in press}

\shorttitle{Chromospheric Evaporation on Proxima Centauri}
\shortauthors{G\"udel et al.}

\begin{document}

\title{X-Ray Evidence for Flare Density Variations and Continual Chromospheric Evaporation
        in Proxima Centauri}

\author{Manuel G\"udel\altaffilmark{1}, 
        Marc Audard\altaffilmark{1,2},
	Stephen L. Skinner\altaffilmark{3},
	Matthias I. Horvath\altaffilmark{4}}
\altaffiltext{1}{Paul Scherrer Institut, W\"urenlingen und Villigen, 
                 CH-5232 Villigen PSI, Switzerland}	
\altaffiltext{2}{Present address: Columbia Astrophysics Laboratory, Mail code 5247,
                 Columbia University,  550 West 120th Street, New York,  NY 10027, USA}	
\altaffiltext{3}{Center for Astrophysics and Space Astronomy, University of Colorado, Boulder CO, 
                 80309-0389, USA}	
\altaffiltext{4}{Institute of Astronomy, ETH Zentrum, CH-8092 Z\"urich, Switzerland}	
	
\begin{abstract}
Using the {\it XMM-Newton} X-ray observatory to monitor the nearest star to
the Sun, Proxima Centauri, we recorded the weakest X-ray flares on a magnetically active 
star ever observed.  Correlated X-ray and optical variability provide strong support 
for coronal energy and mass supply by a nearly continuous sequence of rapid explosive 
energy releases. Variable emission line fluxes were observed in the He-like triplets 
of \ion{O}{7} and \ion{Ne}{9} during a giant flare. They give  direct X-ray evidence for density 
variations, implying densities between $2\times 10^{10} - 4\times 10^{11}$~cm$^{-3}$
and providing estimates of the mass and the volume of the line-emitting
plasma. We discuss the data in the context of the chromospheric evaporation scenario.
\end{abstract}

\keywords{stars: activity --- stars: coronae ---  stars: flare ---
          stars: individual (Proxima Centauri) --- X-rays: stars }

\section{Introduction}

The identification of the physical mechanisms underlying the heating of coronae of magnetically
active stars to some 10--100 Million degrees (MK) remains as one of the fundamental
problems of high-energy stellar astrophysics.
Among the non-steady heating mechanisms, explosive release of
energy by flares has received considerable attention  in solar physics (e.g.,
\citealt{porter95, krucker98, parnell00, aschwanden00}), in particular since 
the statistical ensemble of flares, if extrapolated to small ``micro-''
or ``nano-''\-flares, may provide enough energy to explain the
total coronal energy losses. 

There is mounting evidence that flares also play a fundamental role in coronal
heating of magnetically active stars. The quiescent X-ray
luminosity $L_{\rm X}$ is correlated with the flare rate in X-rays
\citep{audard00} and in the optical \citep{doyle85, skum85}. Statistically,
the hotter X-ray emitting plasma component is more variable than the cooler one,
indicating that flares may be involved \citep{giam96}.  Recent studies
of the flare frequency in magnetically active stars as a function of the flare 
energy indicate power-law distributions that may be  sufficient to explain all  
coronal radiative losses \citep{audard00, kashyap02, guedel03}.

The physical mechanisms that transport the hot plasma into
the corona are, however, less clear. A favored model in solar physics predicts 
that an initial release of high-energy particles (detected by their radio or prompt
optical emission, see below) deposits energy in the cool chromosphere, inducing an 
overpressure that drives hot material into the corona (``chromospheric evaporation''). 
As for any coronal heating mechanism, a large density increase is predicted which 
should be measurable in high-resolution X-ray spectra, in particular in density-sensitive 
He-like triplets of \ion{O}{7} and \ion{Ne}{9}.
 
We present novel observations of Proxima Centauri, the nearest star to the Sun 
(distance = 1.3~pc),  with the {\it XMM-Newton} satellite,  providing
new important evidence  for frequent chromospheric evaporation. Detailed models 
will be presented in a forthcoming paper.
 
\section{Target and Observations}

Proxima Centauri is a dM5.5e dwarf revealing considerable coronal activity. 
Its ``quiescent'' X-ray luminosity $L_{\rm X} \approx 5-10\times 10^{26}$~erg~s$^{-1}$ is
similar to the Sun's  despite its $\approx$50 times smaller surface area. It has 
attracted the attention of most previous X-ray observatories 
\citep{haisch80, haisch81, haisch83, haisch90, haisch95, haisch98, wargelin02}. 

{\it XMM-Newton} \citep{jansen01} observed the star on 2001 August 12 during 65~ks of 
exposure time. For our light curve analysis, we used the most sensitive soft X-ray detector 
system presently available, namely the three combined European Photon Imaging 
Cameras (EPIC; \citealt{turner01, strueder01}) with a  total effective
area of $\approx 2000$~cm$^{2}$ at 1~keV and $\approx 2500$~cm$^{2}$ at 1.5~keV
(with the medium filters inserted). 
The flares thus recorded are the weakest yet observed on an active star. To minimize CCD 
pile-up,  the small window modes were used. Background light curves
were extracted from  CCD fields outside the source region. The Reflection Grating 
Spectrometers  (RGS; \citealt{herder01}) simultaneously recorded high-resolution 
X-ray spectra between 0.35--2.5~keV, with a resolving power of $E/\Delta E \approx 
300$ (FWHM) around the \ion{O}{7} lines at $\approx 22$~\AA. The Optical Monitor  
(OM;  \citealt{mason01})  observed the star through its U band filter in high-time 
resolution mode. All data were analyzed using the {\it XMM-Newton} Science Analysis System 
(SAS versions  5.3 and 5.3.3).

Relevant extracts  of the observations are shown in Figures 1 and 2. 
The first 45~ks of the observation reveal low-level emission around $L_{\rm X} = 
6\times 10^{26}$~erg~s$^{-1}$ (0.15-10~keV). Thanks to the high  sensitivity, 
much of the emission is resolved into a succession of weak X-ray flares (Figure 1). 
The smallest discernible events show $L_{\rm X, 0.15-10} = 
2\times 10^{26}$~erg~s$^{-1}$ and an integrated X-ray energy loss of 
$E_{\rm X, 0.15-10} \approx
1.5\times 10^{28}$~erg, 
corresponding to modest solar flares. Most importantly, many (although not all) of these 
flares are  preceded by a short pulse in the optical U band (Figure 1). A large  flare 
monitored almost in its 
entirety governed the final  20~ks of the observation, with a peak 
$L_{\rm X, 0.15-10} \approx 3.7\times 10^{28}$~erg~s$^{-1}$ and 
$E_{\rm X, 0.15-10} 
\approx 1.5\times 10^{32}$~erg (Figure 2). Due to pile-up in the
MOS detectors, we used only the PN data for the light curve
analysis of this flare. Again, a large optical burst accompanies the X-ray flare
during its increase. The figure also shows spectral extracts 
of the He-like triplets of \ion{O}{7} and \ion{Ne}{9} referring to  different phases of the
large flare. The line flux ratios
are clearly variable, an aspect that will be discussed in detail below.

\begin{figure} 
\epsscale{0.63}
\plotone{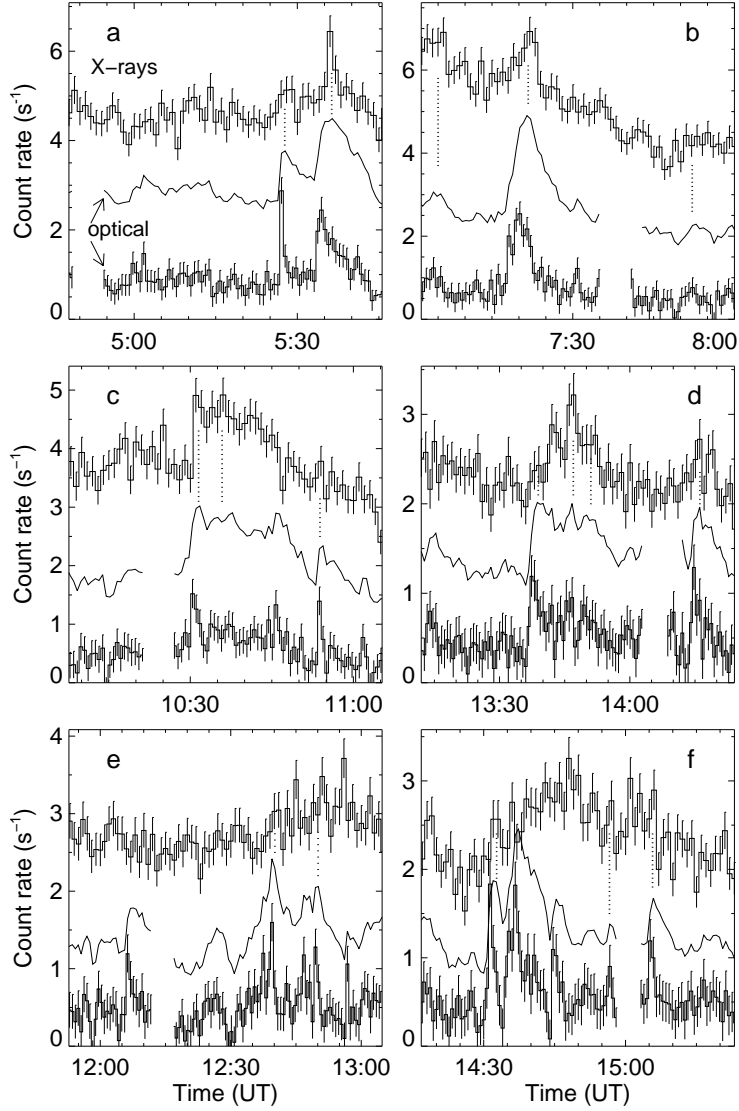}
\caption{Time correlations between X-ray and optical flares. 
The upper curves in each panel show the co-added EPIC X-ray light curve 
 (bin size 60~s, 0.15--4.5~keV). The lower curves illustrate the optical signal,
binned to 40~s for optimum flare detection (and 
scaled in flux by a factor of 0.2 and shifted along the y-axis for 
illustration purposes; the time gaps are of instrumental origin). The thick solid
line is a convolution of the optical light curve with a cut-off exponential
function, and often closely resembles  the X-ray signal (vertical dotted lines mark 
suggestive examples). Panels a and b: Correlated single flares.
Panels c and d:  Groups of flares. Panels e and f: 
Poor detailed correlation, but  optical flare groups occur during general X-ray 
increases.}
\end{figure}

\begin{figure} 
\epsscale{1.}
\plotone{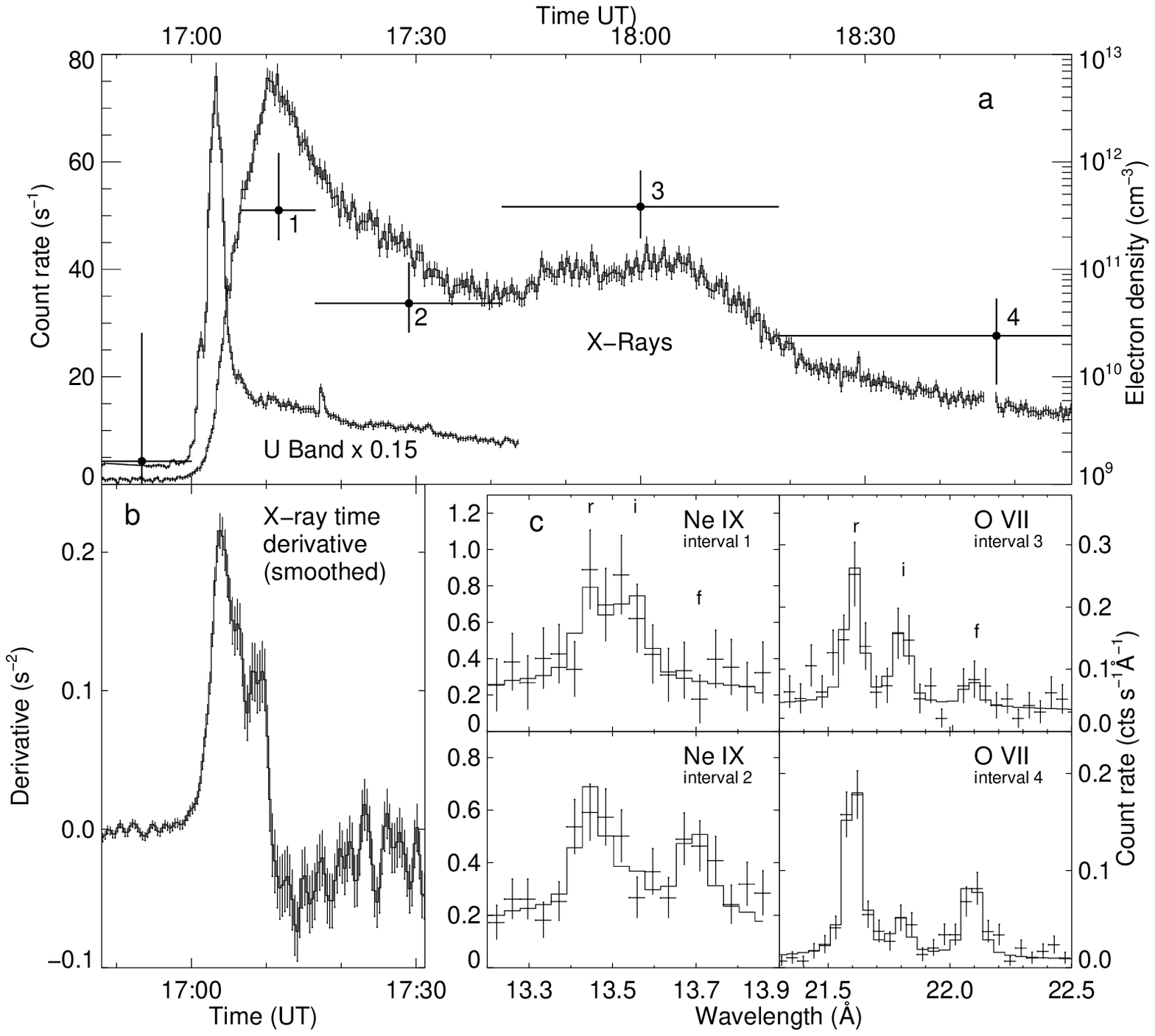}
\caption{Large flare on Proxima Centauri. The upper panel (a) shows 
the  EPIC PN X-ray (0.15--10~keV) and the preceding (scaled) optical light curves (bin size =
20~s).
The crosses refer to the logarithmic axis on the right side and indicate
the electron densities as measured from \ion{O}{7} lines. The arms in
the vertical direction indicate the (linear) 1$\sigma$ error bars. Panel
b is the smoothed time derivative of the X-ray light curve.
It closely resembles the optical signal. Panels c illustrate
He-like triplets with variable $R$ ratios obtained during
the four intervals marked in panel a, two examples for
\ion{O}{7} and two for \ion{Ne}{9}. The histograms
represent fits with three narrow lines each. 
(The \ion{O}{7} triplets of intervals 1 and 2 are very similar to
those of intervals 3 and 4, respectively).}
\end{figure}

\section{Results and Discussion}

We now seek an interpretation within the evaporation scenario.
During a flare, high-energy  electrons are thought to be accelerated in the corona  
from where they travel toward the magnetic footpoints in the  
chromosphere. They induce a prompt signal in the optical U band (3000-4000~\AA), 
probably owing to Balmer emission or blackbody radiation, or both 
\citep{hudson92, neidig93, hawley95}. Their bulk kinetic energy heats the chromospheric 
gas to several MK, building up a pressure gradient that drives  the hot plasma into the corona, 
thus increasing the coronal density and the soft X-ray emission. Since the 
optical signal $L_{\rm O}$ approximately traces the rate of energy deposited by the electrons
\citep{hudson92}, 
and  the X-ray losses $L_{\rm X}$ are a measure of the accumulated coronal
thermal energy, the time  integral of the optical signal should resemble the X-ray light curve,
\begin{equation}\label{eq1} 
L_{\rm X}(t) \propto \int_0^tL_{\rm O}(t^{\prime})dt^{\prime}
\end{equation}
where the integration starts at a point in time preceding the flare ($t^{\prime} = 0$).
In other words, the time derivative of the X-ray light curve should mimic the optical signal:
\begin{equation}\label{eq2}
{dL_{\rm X}\over dt} \propto L_{\rm O}.
\end{equation}
Equation~\ref{eq1} is appropriate for testing faint signals, whereas equation~\ref{eq2} 
requires strong signals but is sensitive to details in the light curves.
This important diagnostic has been observed in  the solar corona, known as  the 
Neupert Effect (using radio or hard X-ray emissions instead  of U band radiation; 
\citealt{neupert68, dennis93}), but has rarely been detected in stellar observations, 
especially in low-level emissions \citep{doyle86, doyle88}, presumably owing to the limited 
sensitivity of previous X-ray devices.  Qualitative evidence is seen in some strong flares 
(e.g., \citealt{dejager86, dejager89, schmitt93, hawley95}). We suggest that each of our optical signals is the 
equivalent   of an impulsive burst, while the associated X-ray flares are the signatures of 
the subsequent chromospheric evaporation. A surprising aspect is the high repetition rate 
of such processes in Proxima Centauri.

While hot plasma is accumulated in the corona,  the X-ray emission of any plasma packet 
decays nearly exponentially due to cooling. Equation 1 in this form does  not consider 
cooling;  we therefore include a decay term by convolving the optical light curve with a  
cut-off exponential decay profile ($e^{-t/\tau}$ for $t \ge 0$ and zero for $t<0$). A decay
time of $\tau = 200$~s is found to be appropriate, and the convolution is shown
by the thick solid curves in Figures 1a-f.  After  the convolution, a high level of 
correlation with the X-rays reveals itself in many coincident peaks and similar light curve shapes
(Figures 1a,b).  Suggestive examples are marked by vertical dotted lines. 
The optical light curve often  predicts 
the subsequent X-ray behavior, revealing an intimate 
connection between the two emissions. The correlation holds even for some
flare groups (Figures 1c,d). In some cases,  the detailed correlation is poor although optical flare 
groups  cluster during  X-ray flux increases (Figures 1e,f).
The large flare allowed us to compute its 
time derivative (Figure 2b). It traces the optical signal closely 
until after the optical flare peak, when uncorrelated cooling mechanisms take over. Here,
$\tau$ was not considered as it appears that ongoing heating on time scales longer than
$\tau$ dominates over cooling.

The X-ray flux $F$ from a collisional plasma is proportional to the volume emission
measure (EM), $F = \mathrm{const}\int n_en_{\rm H}dV \approx  0.84~\mathrm{const}\int n_e^2dV$ where $n_e$ is the 
electron density,  $n_{\rm H}$ is the hydrogen (or proton) density, and $dV$ is the source 
volume element; the proportionality constant is the line emissivity  \citep{mewe85}. The determination
of $V$ and $n_e$ from fluxes alone is therefore a degenerate problem. It can be circumvented 
by using the He-like spectral line triplets as a diagnostic for
$n_e$. The relevant triplets of the \ion{O}{7} and \ion{Ne}{9} ions are each formed by  
radiative decays to the ground state, namely the resonance transition $r$ ($1s^2~^1S_0 - 1s2p~^1P_1$), 
the intercombination transition $i$ ($1s^2~^1S_0 - 1s2p~^3P_{1,2}$), and the forbidden 
transition $f$ ($1s^2~^1S_0 - 1s2s~^3S_1$). In cool stars with negligible 
ultraviolet continuum radiation, the flux ratio $R = f/i$ is sensitive to $n_e$ 
because electron collisions can excite the $^3S_1$ to the $^3P$ state before the former decays radiatively 
\citep{gabriel69, porquet01}. The data reveal strong variations of $R$ along the flare 
(examples in Figure 2c). The  characteristic $n_e$ derived from \ion{O}{7} reaches
$\approx 4\times 10^{11}$~cm$^{-3}$ during the flux maxima, but is much lower  
($\approx 2\times 10^{10}$~cm$^{-3}$) during the decays (crosses in Figure 2a).
Previous attempts to estimate stellar coronal flare densities spectroscopically used lines in the 
extreme ultraviolet range, but density variations were difficult to constrain (e.g., 
\citealt{monsignori96}; for a marginal X-ray result, see \citealt{stelzer02}).
 Qualitatively similar variations are seen in the \ion{Ne}{9} triplet
(Figure 2c), although blends of  the $i$ line thwart a precise quantitative interpretation. 
The strongly variable $f$ line alone suggests that plasma around 4--5~MK (the maximum formation 
temperature of \ion{Ne}{9}) attained densities across the density sensitivity range of this triplet,
$10^{11}-2\times 10^{12}$~cm$^{-3}$ \citep{porquet01}.

To derive the associated EM, we measured  
the total flux $F_{\rm OVII}$ of the  \ion{O}{7} lines used above. We found that $F_{\rm OVII}$ 
is nearly  proportional to the 0.15--10~keV EPIC count rate. This  is important
because \ion{O}{7} lines are efficiently formed only in a temperature range  of 
$1 \la T \la 4$~MK. We  conclude that variations of the EM-weighted formation 
temperature of these lines are not the dominant factor for the 
variability of $F_{\rm OVII}$. The ratio $G = (f+i)/r$ provides an independent test since 
it is a temperature indicator for the \ion{O}{7} line-forming plasma \citep{gabriel69, porquet01}. We infer
characteristic temperatures, $T_G = 1.5-3.5$~MK, that cluster around the maximum  formation 
temperature ($T_{\rm OVII}\approx 2$~MK) but do not show a clear correlation with $F_{\rm OVII}$. We will 
therefore approximate the \ion{O}{7} emitting plasma as being isothermal at $T_{\rm OVII}$;   
systematic errors thus introduced for the EM are no larger than a factor of 
$\approx$2 (see tables in \citealt{mewe85}). For a homogeneous source at $T_{\rm OVII}$,
$F_{\rm OVII}  \propto n_e^2V$. Therefore,
$V \propto F_{\rm OVII}/n_e^2$,  and the total mass $M$ involved 
is $M \propto F_{\rm OVII}/n_e$. We used emissivities from \citet{mewe85} 
referring to solar photospheric composition. From spectral fits in
XSPEC \citep{arnaud96}, we 
found an abundance of oxygen of $0.52\pm 0.08$ times the solar 
photospheric value \citep{anders89}. For the primary peak, 
its decay, the secondary peak, and its decay, the  masses of \ion{O}{7} emitting plasma thus amount to 
$M\approx 4.8\times 10^{14}$~g, $2.1\times 10^{15}$~g, $3.2\times 10^{14}$~g, and $2.5\times 10^{15}$~g, 
respectively, while $V \approx 7.0\times 10^{26}$~cm$^3$, $2.3\times 10^{28}$~cm$^3$, 
$4.3\times 10^{26}$~cm$^3$, and $5.3\times 10^{28}$~cm$^3$, respectively. The potential energy
of a $10^{15}$~g plasma packet at a height of $9\times 10^3$~km
(the pressure scale height at $2\times 10^6$~K for a radius of $10^{10}$~cm and a stellar mass
of 0.2$M_{\odot}$, see \citealt{haisch83} and references therein) is $2.5\times 10^{29}$~erg.
This compares with the total thermal energy content of the same plasma, 
$E \approx 1.65\times (3/2)MkT/m_{\rm H} \approx 4\times 10^{29}$~erg, while the total radiated 
X-ray energy from  the cooler plasma component at 0.2 keV (determined from a spectral fit)
is approximately $1.7\times 10^{31}$~erg during the complete flare (using a peak EM of
$1\times 10^{50}$~cm$^{-3}$, a decay time of 4300~s, and a cooling loss rate of 
$4\times 10^{-23}$~erg~s$^{-1}$cm$^3$), necessitating considerable replenishment of the
cool material during the flare.

Even after consideration of the uncertainties  due to variations of $T_G$, it
is clear that densities, masses and volumes of the plasma accessible by 
\ion{O}{7} increase during  the flare. Coronal density increases are a consequence
of any heating mechanism in closed loops. It is the preceding optical impulsive
emission that strongly supports chromospheric
evaporation induced by electron beams. Solar white-light flares are closely 
correlated in time with hard X-ray bursts, suggesting that this phase relates
to electron bombardment of the chromosphere \citep{hudson92, neidig93}, while soft
X-rays and higher densities are evolving more gradually.
Since plasma cooling becomes important early in the flare development, the relatively
cool \ion{O}{7} emitting plasma component is not only augmented due to heating of 
chromospheric material, but  also due to cooling of hotter material that has initially 
reached temperatures beyond the formation range of \ion{O}{7} lines, hence the continuous
increase of masses and volumes during the decay.

The low flux of the weaker flares and their strong mutual overlap prohibit individual 
density measurements, but integration of larger segments shows appreciable changes in 
$R$ as well, with the highest-density episode related to the segment in Figure 1b (an 
increase of at most a factor of ten over the average). Furthermore, the time behavior of 
several optical and X-ray flares that expresses itself as a Neupert Effect is similar to 
the behavior seen in the large flare, indicating that the same  processes repeat at a much 
higher cadence at lower energy levels such that they govern the X-ray emission  most 
of the time. We do not find a one-to-one correspondence between all optical and X-ray flares,
but this is no different on the Sun \citep{dennis93}. 
Some groups of small optical flares occur, however, during  a general increase of the 
soft X-ray light curve. 

The present observations suggest an important role of chromospheric evaporation
in active stellar coronae. The resulting high plasma densities appear to be a
critical element for the high X-ray luminosities of magnetically active stars.

\acknowledgments  Helpful comments by the referee are acknowledged. Research at PSI  has been supported by the Swiss National 
                  Science Foundation (grant 2000-058827). 
		  The present project is based 
                  on observations obtained with {\it XMM-Newton}, an ESA science 
                  mission with instruments and contributions directly funded by 
                  ESA Member States and the USA (NASA).  

%\clearpage
% References: If > 8 then only first et al.


\begin{thebibliography}{}
\bibitem[Anders \& Grevesse(1989)]{anders89}Anders, E., \& Grevesse, N. 1989,
         Geochim.  Cosmochim. Acta, 53, 197
\bibitem[Arnaud(1996)]{arnaud96}Arnaud, K.~A. 1996,  in Astronomical
         Data Analysis Software and Systems V, ed. G. Jacoby \& J. Barnes 
        (San Francisco: ASP), 17
\bibitem[{Asch}\-wan\-den et al.(2000)]{aschwanden00} Aschwanden, M.~J., Tarbell,
         T.~D., Nightingale, R.~W., Schrijver, C.~J., Title, A., Kankelborg, C.~C.,
         Martens, P.~C.~H., \& Warren, H.~P. 2000, \apj, 535, 1047
\bibitem[Audard et al.(2000)]{audard00}Audard, M.,  G\"udel,  M.,
         Drake, J.~J., \&  Kashyap, V.~L. 2000, ApJ, 541, 396
\bibitem[de Jager et al.(1986)]{dejager86}de Jager, C., et al. 1986,
        A\&A, 156, 95
\bibitem[de Jager et al.(1989)]{dejager89}de Jager, C.,  et al. 1989,
        A\&A, 211, 157
\bibitem[den Herder et al.(2001)]{herder01}den Herder, J.~W., et al. 2001,
        A\&A, 365, L7
\bibitem[Dennis \& Zarro(1993)]{dennis93}Dennis, B.~R., \& Zarro, D.~M. 1993,   
        Solar Phys., 146, 177
\bibitem[Doyle \& Butler(1985)]{doyle85}Doyle, J.~G., \& Butler, C.~J.
         1985, Nature, 313, 378
\bibitem[Doyle et al.(1988)]{doyle88}Doyle, J.~G., Butler, C.~J., Byrne, P.~B.,
         \& van den Oord, G.~H.~J. 1988, A\&A, 193, 229
\bibitem[Doyle et al.(1986)]{doyle86}Doyle, J.~G.,  Butler, C.~J., Haisch, B.~M.,
        \& Rodon\`o, M. 1986, MNRAS, 223, 1P 
\bibitem[Gabriel \& Jordan(1969)]{gabriel69}Gabriel, A.~H., \& Jordan, C. 1969, 
        MNRAS, 145, 241   
\bibitem[Giampapa et al.(1996)]{giam96} Giampapa, M.~S., Rosner, R.,
         Kashyap, V., Fleming, T.~A., Schmitt, J.~H.~M.~M., \& Bookbinder, J.~A.  
	 1996, ApJ,  463, 707
\bibitem[G\"udel et al.(2003)]{guedel03}G\"udel, M.,  Audard, M.,
         Kashyap, V.~L., Drake, J.~J., \& Guinan, E.~F. 2003,  ApJ, in press
\bibitem[Haisch et al.(1995)Hiasch, Antunes, \& Schmitt]{haisch95}Haisch, 
        B.~M.,  Antunes, A., \& Schmitt, J.~H.~M.~M. 1995, Science,  268, 1327
\bibitem[Haisch et al.(1990)]{haisch90}Haisch, B.M., Butler, C.J., Foing, B.,
         Rodon\`o, M.,  \&  Giampapa, M.S. 1990, A\&A, 232, 387
\bibitem[Haisch et al.(1980)]{haisch80}Haisch, B.~M., Harnden, F.~R. Jr.,
         Seward, F.~D., Vaiana, G.~S., Linsky, J.~L., \&  Rosner, R. 1980,
        ApJ, 242, L99  
\bibitem[Haisch et al.(1998)]{haisch98}Haisch, B.~M., Kashyap, V.,
         Drake, J.~J.,  Freeman, P. 1998, A\&A, 335, L101
\bibitem[Haisch et al.(1983)]{haisch83}Haisch, B.~M., Linsky, J.~L.,
         Bornmann, P.~L., Stencel, R.~E., Antiochos, S.~K., Golub, L., \&
         Vaiana, G. S. 1983, ApJ, 267, 280 
\bibitem[Haisch et al.(1981)]{haisch81}Haisch, B.~M., et al. 1981, 
        ApJ, 245, 1009  
\bibitem[Hawley et al.(1995)]{hawley95}Hawley, S.~L., et al. 1995, 
        ApJ, 453, 464
\bibitem[Hudson et al.(1992)]{hudson92}Hudson, H.~S., Acton, L.~W., Hirayama, T.,
         \& Uchida, Y. 1992, PASJ, 44, L77
\bibitem[Jansen et al.(2001)]{jansen01}Jansen, F., et al. 2001,
        A\&A, 365, L1
\bibitem[Kashyap et al.(2002)]{kashyap02}Kashyap, V.~L., Drake, J.~J.,
         G\"udel, M., \&  Audard, M. 2002,  ApJ, in press
\bibitem[Krucker \& Benz(1998)]{krucker98}Krucker, S., \& Benz, A.~O. 
         1998, ApJ, 501, L213.
\bibitem[Mason et al.(2001)]{mason01}Mason, K., et al. 2001 
        A\&A, 36, L36
\bibitem[Mewe et al.(1985)Mewe, Gronenschild, \& van den Oord]{mewe85} 
        Mewe, R., Gronenschild,E.~H.~B.~M., \& van den Oord, G.~H.~J. 1985, 
        A\&AS, 62, 197
\bibitem[Monsignori-Fossi et al.(1996)]{monsignori96}Monsignori-Fossi, B.~C.,
         Landini, M., del Zanna, G., \& Bowyer, S. 1996, ApJ, 466, 427 
\bibitem[Neidig \& Kane(1993)]{neidig93}Neidig, D.~F., \& Kane, S.~R.
        1993, Solar Phys., 143, 201
\bibitem[Neupert(1968)]{neupert68}Neupert, W.~M. 1968,
	ApJ, 153, L59
\bibitem[Parnell \& Jupp(2000)]{parnell00} Parnell, C.~E., \& Jupp, P.~E.
         2000, \apj, 529, 554
\bibitem[Porquet et al.(2001)]{porquet01}Porquet, D., Mewe, R.,  Dubau, J.,
        Raassen, A.~J.~J., \& Kaastra, J.~S. 2001,
        A\&A, 376, 1113 
\bibitem[Porter et al.(1995)Porter, Fontenla, \& Simnett]{porter95} Porter,
         J.~G., Fontenla, J.~M., \& Simnett, G.~M. 1995, \apj, 438, 472
\bibitem[Schmitt et al.(1993)]{schmitt93}Schmitt, J.~H.~M.~M., Haisch, B., \& Barwig, H. 1993,
         ApJ, 419, L81 
\bibitem[Skumanich(1985)]{skum85} Skumanich, A. 1985, Aust.~J.~Phys., 38, 971
\bibitem[Stelzer et al.(2002)]{stelzer02}Stelzer, B.,  et al. 2002,
         A\&A, 392, 585
\bibitem[Str\"uder et al.(2001)]{strueder01}Str\"uder, L., et al. 2001, 
        A\&A, 365, L18
\bibitem[Turner et al.(2001)]{turner01}Turner, M.~J.~L., et al. 2001,  
        A\&A, 365, L27
\bibitem[Wargelin \& Drake(2002)]{wargelin02}Wargelin, B.~J., \& Drake, J.~J. 2002,  
        ApJ, 578, 503
\end{thebibliography}
\end{document}